# Magnetism, transport and atomic structure of amorphous binary $Y_xCo_{1-x}$ alloys.

Zexiang Hu, Jean Besbas, Katarzyna Siewierska, Ross Smith, Plamen Stamenov and  J. M. D. Coey*

School of Physics and CRANN, Trinity College, Dublin 2 Ireland.

Abstract

Sputtered thin films of binary $Y_xCo_{1-x}$ with $0 < x \leq 0.54$ and thickness $\approx 15$ nm are investigated to help understand the ferromagnetism of cobalt in amorphous rare-earth cobalt alloys. The magnetic moment per cobalt falls to zero at $x_0 \approx 0.50$, where the appearance of magnetism is marked by a para-process with a dimensionless susceptibility of up to 0.015. All films are magnetically soft, with densities that fall between those of crystalline Y-Co intermetallic compounds and the density of a relaxed 10,000-atom binary random close-packed model of hard spheres with an Y:Co volume ratio of 3:1, where the packing fractions for all films lies in a narrow range $0.633 \pm 0.004$ and Co is coordinated by 3.2 Co and 3.2 Y atoms at $x = 0.5$. All films with $x < 0.4$ exhibit in-plane shape anisotropy that is about six times as great as an intrinsic perpendicular component. Average cobalt spin and orbital moments obtained by X-ray magnetic circular dichroism were 1.31 $\mu_B$ and 0.32 $\mu_B$, respectively for amorphous $Y_{0.25}Co_{0.75}$. Strong local anisotropy is associated with the large cobalt orbital moment, but there is little influence of anisotropy on the ferromagnetic order because of exchange averaging. The Hall effect and magnetoresistance are modelled in terms of effective uniform rotation of the magnetization, with spontaneous and band contributions. Amorphous $Y_xCo_{1-x}$ is contrasted with amorphous $Y_xFe_{1-x}$, which exhibits random noncollinear magnetic order that is very sensitive to the film density.

*jcoey@tcd.ie

# 1 Introduction

The magnetism of rare-earth (R) transition-metal (T) intermetallic compounds with T = Fe, Co, Ni was studied intensely from the 1960s to the 1990s [1, 2]. An understanding emerged of the main interactions influencing the magnetic order — exchange among and between the magnetic moments of atoms on the transition metal and rare-earth sublattices, and the crystal field interaction of the rare-earth electric quadrupole moments with electrons of surrounding atoms, which is the major source of magnetic anisotropy in many of these compounds. An outcome of the early work was discovery and development of the new rare-earth permanent magnets, including $SmCo_5$, $Sm_2Co_{17}$, $Nd_2Fe_{14}B$ and $Sm_2Fe_{17}N_3$ [3, 4].

Around the same time, reliable methods for fabricating high-quality magnetic thin films of both crystalline and amorphous metals and alloys were being developed, which led to the emergence of spin electronics. Questions of how magnetic order could exist in the absence of a crystalline lattice, and the nature of the atomic and magnetic structure of amorphous metals were research preoccupations. As investigations extended from crystalline to amorphous binary rare-earth transition-metal alloys, new noncollinear magnetic structures associated with antiferromagnetic T–T exchange, or random 'crystal field' anisotropy at the R sites were discovered [5,6]. When R is Gd and T–T exchange is predominantly ferromagnetic, as it is in R-Co alloys, the alloys are collinear ferrimagnets where the Gd atoms couple antiparallel to their Co neighbours, but when R is a heavier rare-earth atom, the anisotropy may lead to noncollinear ferrimagnetic (sperimagnetic) structures [7], at least at low temperature. Amorphous ferrimagnetic Gd-(Fe,Co) thin films became the mainstay of a method of magneto-optic recording based on Curie point or compensation point writing [8], which later became obsolete because the minimum spot size for writing was limited by the wavelength of light. A revival of interest in ferrimagnetic a-Gd-(Fe,Co) alloys (we use 'a-' to denote an amorphous alloy) that followed the discovery of ultra-fast all-optical magnetic switching induced by picosecond laser pulses [9, 10] reopened some questions that were imperfectly resolved in the last century regarding the nature of the binary atomic structures and the noncollinear magnetism of a-R–T alloys.

Our focus here is on the binary magnetic alloy system, a-$Y_xCo_{1-x}$. Yttrium is an ideal nonmagnetic proxy for the magnetic rare earths from Gd to Tm in the second half of the $4f$ series. It is a $4d$ element with the same atomic radius as gadolinium (180 pm), but no $4f$ electrons. Its $4d$ and $5s$ electrons occupy similar orbitals to the $5d$ and $6s$ electrons of the main rare-earth series. Yttrium itself is a Pauli paramagnet with susceptibility $1.2 \times 10^{-5}$. Crystalline cobalt is a strong ferromagnet with a full majority-spin $3d$ sub-band and a spin moment of 1.6 $\mu_B$ corresponding to the number of holes in the minority-spin $3d$ sub-band; there is also a small orbital moment of 0.14 $\mu_B$ and the Curie temperature is 1360 K. The Co atomic radius is 125 pm, and its atomic volume is a third of that of Y. Because of their disparate sizes, yttrium and cobalt form a series of crystalline intermetallic compounds ranging from $Y_2Co_{17}$ to $Y_3Co$. However, it is possible to vary the composition continuously from almost pure cobalt Co to almost pure Y in rapidly quenched binary amorphous thin films, free from the constraint of an ordered crystal structure. The magnetism of a-$Y_xCo_{1-x}$ is a stepping stone from which to evaluate the more complex magnetism of other a-$R_xCo_{1-x}$ alloys with a magnetic rare earth.

Early work on the a-$Y_xCo_{1-x}$ system addressed the magnetization and appearance of magnetism [11, 12], local moment formation [13, 14], Curie temperature [14], anisotropy [15] and the effects of sputtering pressure [16] and oxidation [17] on the thin film structure. Topics of interest in the present work are the orbital moment of cobalt, the magnetic anisotropy of the thin films, transport properties of the Co subnetwork and the structural condition for the appearance of a cobalt moment.

## 2    Thin Films

Films of $Y_xCo_{1-x}$ approximately 15 nm thick with 18 compositions ranging from $x = 0$ to $x = 0.55$ were grown on oxidized silicon wafers by DC sputtering from separate 50 mm yttrium and cobalt targets in a Shamrock sputtering system with a base pressure of about $8 \times 10^{-8}$ torr. No metallic underlayer was used. Films were capped with a 2–3 nm thick layer of $SiO_2$ or $Al_2O_3$ to protect them from oxidation. Film composition was determined from the calibrated deposition rates of the individual targets.

The films were characterized by X-ray diffraction and small-angle X-ray scattering using a Philips Panalytical X'pert Pro diffractometer. No Bragg reflections from any Y–Co intermetallic phase were seen in the diffraction patterns. Only the $x = 0$ cobalt end member showed a broadened diffraction pattern from nanocrystalline cubic close-packed cobalt with a crystallite size estimated as 8.6 nm. The densities and roughnesses of the film and cap layers as well as their thicknesses were fitted using the X'pert Reflectivity Pro software. Some typical data shown in Fig. 1, include the thickness distributions showing peaks for the film and film + cap. The measured densities in g/cm³ are plotted in Fig 2, together with the densities of Y, Co and nine crystalline $Y_nCo_m$ intermetallics, as well as the density of a 10,000 atom relaxed Y-Co binary atomic model discussed in section 4. The experimental packing fractions all lie above the 0.63 value for random close packing [18], but tend towards 0.74 as the cobalt atoms become a smaller fraction of the total and begin to occupy small interstices in the random close-packed yttrium subnetwork.

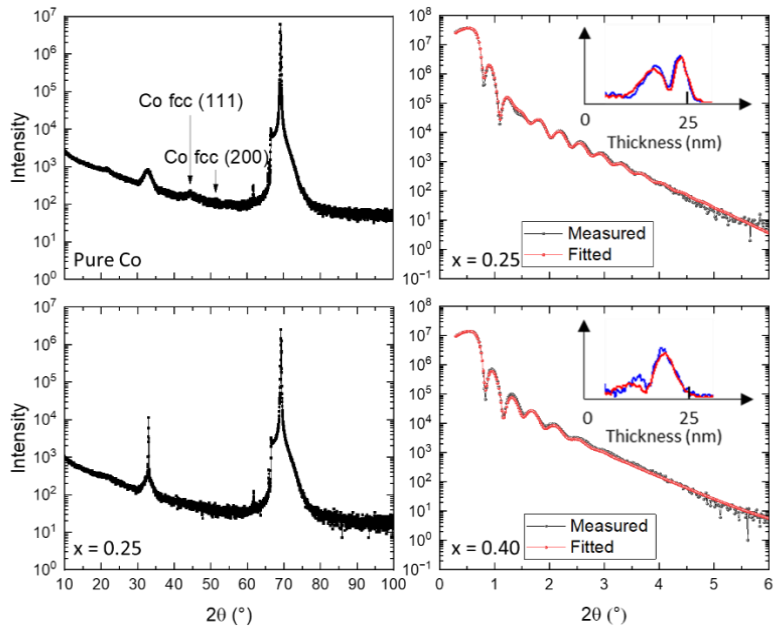

Figure 1. X-Ray diffraction patterns of thin films with $x = 0$ and $x = 0.25$ (left). Small angle X-Ray scattering of films with thickness $t$, $x = 0.25$ and $x = 0.40$ (right). The red lines are fits to the data.

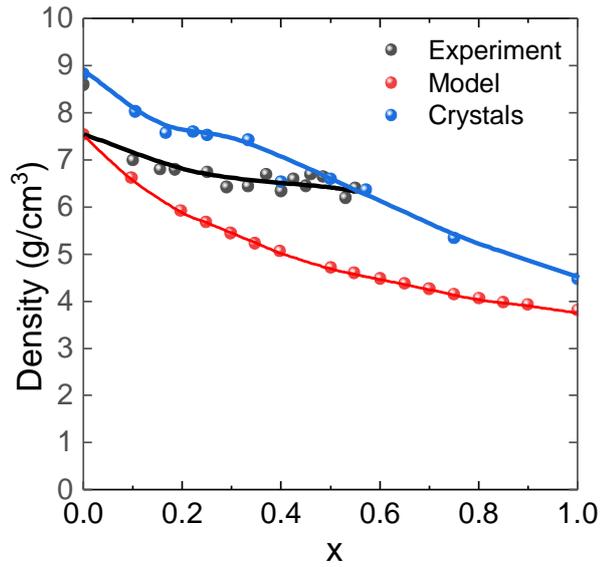

Figure 2. Density of crystalline intermetallic Y-Co compounds (blue), amorphous $Y_xCo_{1-x}$ thin films (black) and the relaxed 10,000 atom amorphous Y-Co model (red).

Table 1. Structural and magnetic data on amorphous $Y_xCo_{1-x}$ thin films.

| x | Main layer thickness (nm) | Main layer roughness (nm) | Main layer density (g/cm3) | Capping layer | Capping layer thickness (nm) | $M_s$ (kA/m) | $m$ ($\mu_B$/Co) | $H_s$ (kA/m) |
|---|---|---|---|---|---|---|---|---|
| 0 | 21.6 | 1.6 | 8.6 | SiO2 | 4.6 | 1431.3 | 1.80 | 1256.5 |
| 0.10 | 16.5 | 1.2 | 7.0 | SiO2 | 4.1 | 986.8 | 1.76 | 737.7 |
| 0.155 | 19.5 | 1.5 | 6.8 | Al2O3 | 3.2 | 911.6 | 1.88 | 736.1 |
| 0.185 | 19.8 | 1.2 | 6.8 | SiO2 | 3.1 | 823.3 | 1.72 | 668.5 |
| 0.25 | 19.7 | 1.3 | 6.8 | SiO2 | 3.2 | 568.5 | 1.35 | 450.4 |
| 0.25 | 8.8 | 1.0 | 6.5 | SiO2 | 2.0 | 705.4 | 1.72 | 647.0 |
| 0.25 | 19.0 | 1.3 | 6.5 | SiO2 | 2.0 | 721.7 | 1.76 | 598.4 |
| 0.29 | 20.9 | 1.8 | 6.4 | Al2O3 | 2.9 | 531.3 | 1.43 | 428.9 |
| 0.333 | 19.6 | 1.8 | 6.4 | Al2O3 | 2.9 | 449.4 | 1.35 | 319.9 |
| 0.37 | 17.5 | 1.5 | 6.7 | SiO2 | 3.1 | 340.5 | 1.04 | 265.8 |
| 0.40 | 17.6 | 1.7 | 6.4 | SiO2 | 3.5 | 328.6 | 1.12 | 163.1 |
| 0.40 | 16.5 | 1.9 | 6.3 | Al2O3 | 3.1 | 241.5 | 0.86 | 150.4 |
| 0.425 | 17.7 | 1.9 | 6.6 | SiO2 | 3.6 | 304.5 | 1.09 | 180.6 |
| 0.45 | 15.6 | 1.7 | 6.7 | SiO2 | 2.4 | 163.2 | 0.63 | 113.0 |
| 0.45 | 15.3 | 2.0 | 6.2 | Al2O3 | 3.1 | 58.3 | 0.26 | nd |
| 0.46 | 14.8 | 1.4 | 6.7 | SiO2 | 3.8 | 118.7 | 0.49 | nd |
| 0.485 | 14.3 | 1.4 | 6.7 | SiO2 | 3.9 | 40.5 | 0.14 | nd |
| 0.53 | 14.5 | 1.3 | 6.2 | SiO2 | 4.1 | 15.6 | 0.08 | nd |
| 0.54 | 14.1 | 1.4 | 6.8 | SiO2 | 3.8 | 5.8 | 0.05 | pp |
| 0.55 | 14.4 | 1.2 | 6.4 | SiO2 | 4.1 | 6.8 | 0.01 | pp |

nd  Not determined.  pp Para process.

The structural properties of the 20 thin films are summarized in Table 1.

## Magnetic Properties

Magnetic moments were measured in a 5 T Quantum Design SQUID magnetometer with the film mounted parallel (IP) or perpendicular (OOP) to the vertical field direction. Small geometrical corrections were applied to correct for the substrate dimensions in the two positions [19], and the diamagnetic susceptibility of the silicon substrate was subtracted to obtain the magnetic moment $m$ of the sample as a function of applied field. The magnetization $M$ in Am$^{-1}$ is calculated from the sample volume, and the moment in Bohr magnetons ($\mu_B$) per cobalt is deduced using the molecular weight and the density. Data are presented in Table 1. Thin film magnetization curves for four different compositions are illustrated in Fig. 3. They show that the easy direction of magnetization lies in-plane for all the amorphous films that have a clear spontaneous magnetization. The average ferromagnetic moment per cobalt atom is plotted for the amorphous alloys in Fig 4a) and for the crystalline intermetallic compounds in Fig 4b). All films are magnetically soft, with a coercivity less than ≈ 1 mT

A spontaneous magnetic moment appears below a different critical concentration $x_0$ for crystalline and amorphous alloys. In the crystalline state, $x_0 \lesssim 0.33$; YCo$_2$ is a band metamagnet with an enhanced Pauli susceptibility of $0.8 \times 10^{-3}$ and a first-order transition to ferromagnetism in 70 T [20]. There is a big difference in the amorphous state, where $x_0 \approx 0.50$. Practically the same value is deduced from room-temperature data or from the magnetization extrapolated to $T = 0$. Curie temperatures when $x < 0.4$ are sufficiently high for the low temperature correction to be negligible [34]. The magnetization exhibits a para process with a broad maximum in the high-field susceptibility in the vicinity of $x_0$, which is illustrated in Fig. 4c). The para process is characterized by an isotropic nonlinear magnetization curve passing through the origin. Close to $x_0$ it is superposed on a small in-plane ferromagnetic moment. There is no coercivity. These samples are marked 'nd' in last column of Table 1.

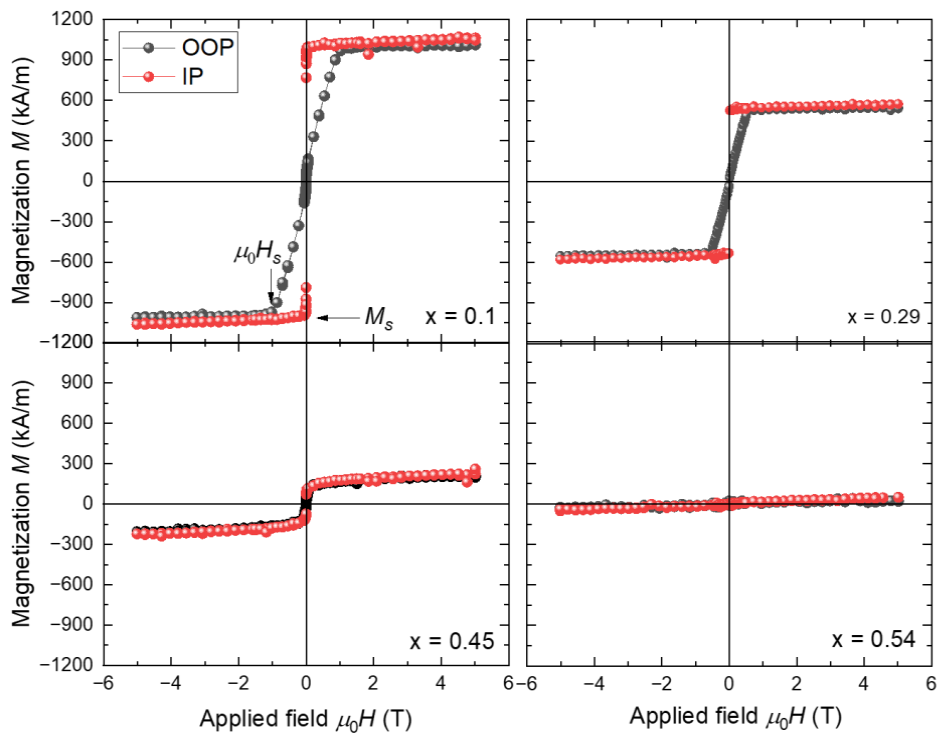

Figure 3. Room-temperature magnetization curves for amorphous thin films of $Y_xCo_{1-x}$ with $x$ = 0.1, 0.29, 0.45 and 0.54.

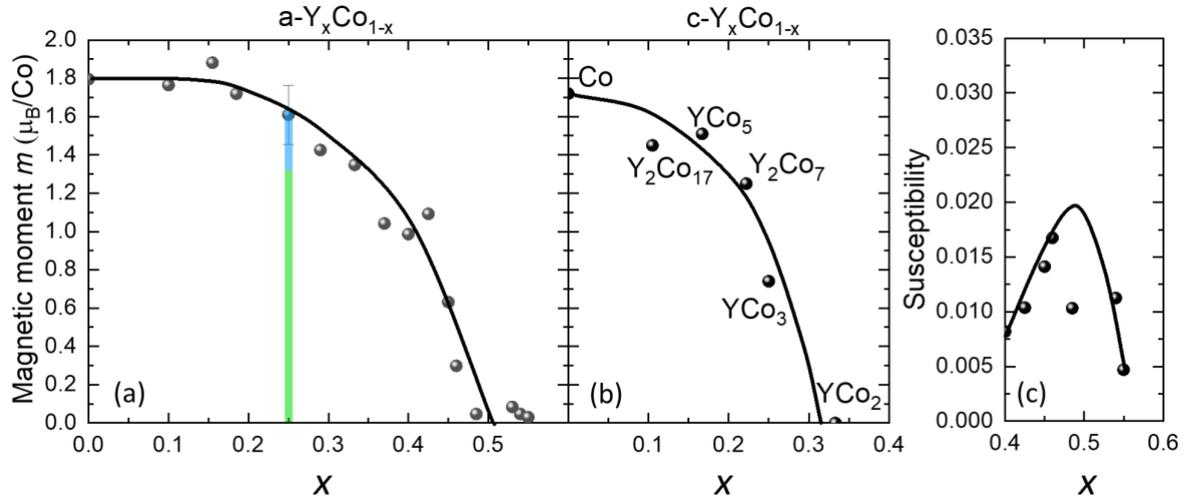

Figure 4. The average magnetic moment per cobalt in a) amorphous $Y_xCo_{1-x}$ alloys and b) crystalline $Y_nCo_m$ intermetallic compounds. The vertical bar in a) represents the spin moment (green) and orbital moment (blue) determined by XMCD. c) is a plot of the dimensionless high-field susceptibility in the vicinity of $x_0$, Solid black lines are guides to the eye.

A film of a-$Y_{0.25}Co_{0.75}$ was studied by X-ray magnetic circular dichroism (XMCD) on the UE49 SGM beamline of the BESY II light source at Helmholtz-Zentrum Berlin-Adlershof. Circularly polarised X-ray absorption spectra (XAS) at normal incidence were measured at 300 K by total electron yield via the drain current at the Co $L_{2,3}$ absorption edge. The film was mounted on a permanent magnet providing a field sufficient to saturate the moment out-of-plane. A well-defined difference spectrum obtained for left and right circularly-polarized radiation, is shown in Fig. 5. Data were analysed using the sum rules [21] to obtain spin and orbital moments of 1.31 $\mu_B$/Co and 0.32 $\mu_B$/Co, respectively. The average total moment is 1.63 $\mu_B$, compared with the value of 1.61 ± 0.18 in Fig 4a), where the error is the standard deviation of the mean of three different samples.

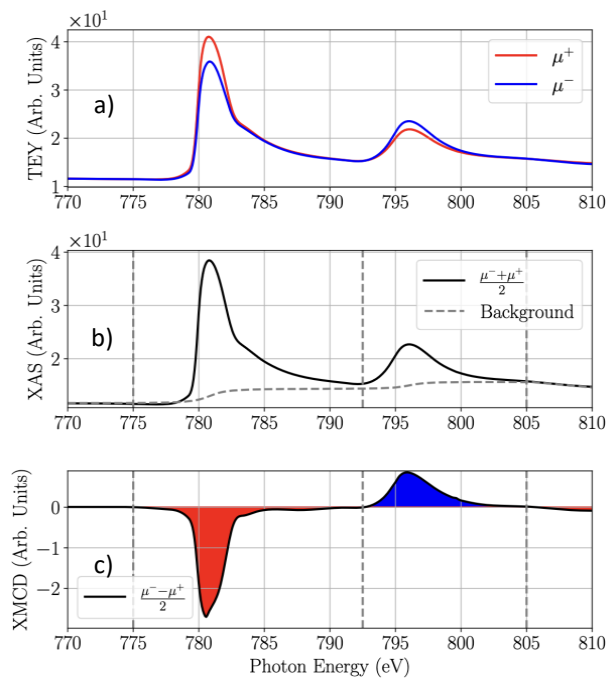

Figure 5. a) The polarization-dependent Co L-edge X-ray absorption spectra (XAS) b) the average XAS with background fit c) XMCD spectrum for a magnetically saturated a-$Y_{0.25}Co_{0.75}$ film.

## 4 Transport Properties.

Resistance, magnetoresistance and Hall effect were measured at room temperature for a selection of thin films using the van der Pauw method with a current of 1mA. A variable magnetic field of up to ±2 T was applied using an electromagnet. Except for the nanocrystalline cobalt film, $x = 0$, which has $\rho_{xx} = 30$ µΩcm, the resistivity of all the amorphous films was in the range 260 – 800 µΩcm with little temperature dependence. Fig. 6 shows the transverse magnetoresistance $\Delta\rho_{xy}/\rho_{xx}$ and Hall effect for four different compositions. In each case, the Hall data as a function of perpendicular magnetic field are separated into an asymmetric part representing the superposition of the anomalous and normal components of the Hall effect, plotted in red, and a symmetric part representing the transverse magnetoresistance, plotted in black, which is the superposition of a band component varying as $B^2$ and a saturating magnetization-related component varying as $\mu_0 M_z^2$. The sign of the magnetoresistance can appear positive or negative on account of offsets of the equipotentials with respect the Hall contacts and a corresponding nonzero component of current along the y-axis. For the samples illustrated the leakage current through the silicon substrate due to defects at the cleavage faces of the chips amounts to no more than 20% of the x-axis current.

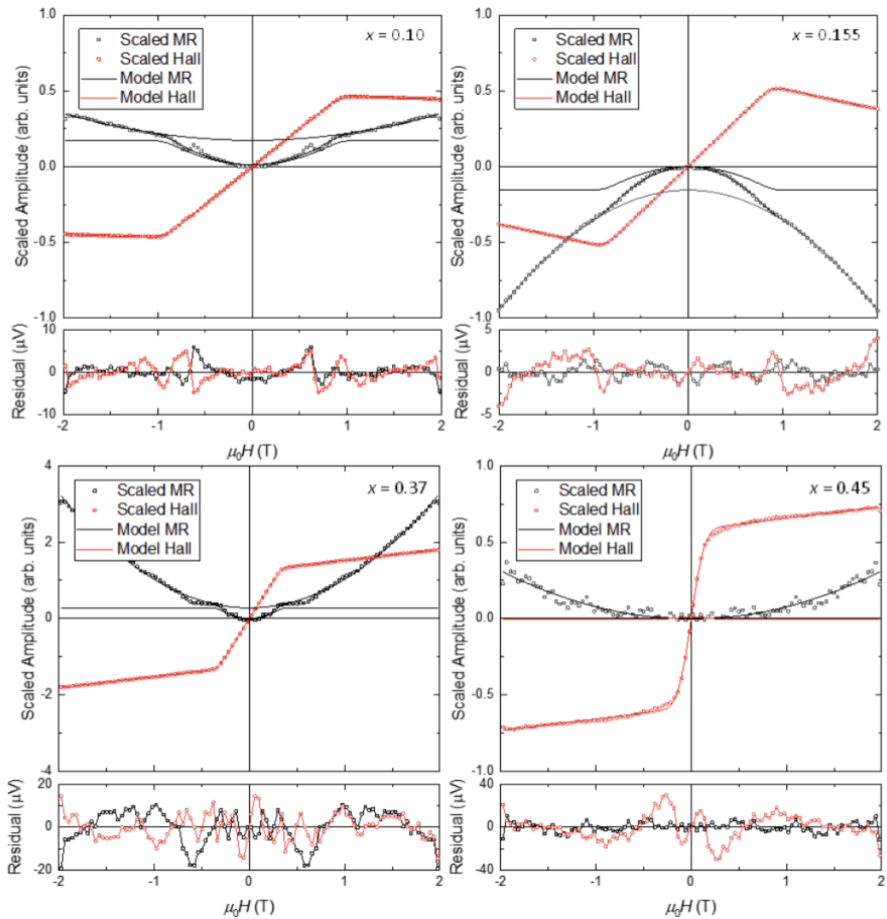

Figure 6. Magnetoresistance and Hall effect of four a-$Y_xCo_{1-x}$ thin films, measured at room temperature. The symmetric (black) and antisymmetric (red) components in the large panels represent the transverse magnetoresistance. and the normal and anomalous Hall effects. The residuals from simultaneous fits to a model of homogeneous rotation of the magnetization (see Section 7) are shown below each panel.

## 5  Binary Amorphous Structure

We modelled the atomic structure of the a-$Y_xCo_{1-x}$ alloys following a procedure introduced by Clark and Wiley to generate random close-packed binary alloys [22]. A total of 10,000 spheres representing Y and Co atoms were picked at random with probabilities $x$ and $(1 – x)$, and positioned in a cube with periodic boundary conditions. A Y:Co volume ratio of 3:1 was chosen for spheres to correspond to the metallic radii of Y and Co atoms. At each step, every atom was moved to reduce overlap with its neighbors. Convergence of the packing fraction resulted from periodically increasing and decreasing the radii of all the spheres, while maintaining the radius ratio. Large radii maximise overlap and displacement of the atoms. Small radii allow atoms to fit into small interstices. Small random displacements were also used to eliminate jamming, allowing the model to find higher-density configurations. Remarkably, the packing fraction converges to the value for random dense packing, not only for the end members $x = 0$ and $x = 1$ but also over the entire range of $x$ to within 1 %. It fluctuated in the range from 0.629 – 0.637, in agreement with the results of Clark and Wiley [22]. Information on the number of nearest-neighbors was obtained from the resulting set of atomic positions. Counting the number of atoms of each species surrounding every atom, we found four distributions representing the number of B nearest-neighbors surrounding an A atom where A and B is either be Y or Co. The distributions were gaussian and a fit was used to extract their average values plotted on Fig. 7.

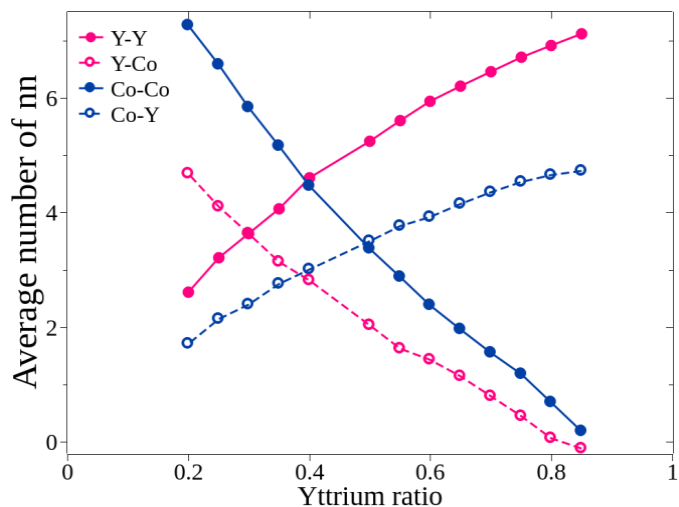

Figure 7. Average numbers of Y and Co neighbours around an Y atom, and Y and Co neighbours around a Co atom in the binary random dense-packed amorphous model.

In the a-$Y_{0.25}Co_{0.75}$ alloys, the average Y:Co coordination of a cobalt atom is 6.6:2.0, while at $x_0 = 0.5$, the critical concentration for the appearance of a cobalt moment, the average Y:Co coordination of a cobalt atom is 3.2:3.2. As in crystalline, Laves-phase $YCo_2$, three Y neighbours are enough to destroy the cobalt moment.

## 6 Discussion

All our thin films of a-$Y_xCo_{1-x}$ with $x \leq 0.4$ exhibit in-plane magnetization as a result of shape anisotropy (Fig. 3). If this were the only contribution, the saturation field $H_s$ in the hard direction perpendicular to the film would be equal to the saturation magnetization $M_s$ of the film, but $H_s$ is actually consistently less than $M_s$. The slope of the graph in Fig. 8 is 1.18, showing that there is an intrinsic perpendicular component that is 15% of the shape anisotropy. When a magnetic field $\mathbf{H}$ is applied perpendicular to film and the net magnetization is represented by a vector $\mathbf{M}$ that rotates coherently and makes an angle $\theta$ with the normal to the film, the magnetostatic energy $E$ contains three terms:

$$E(\theta) = \tfrac{1}{2}\mu_0(M\cos\theta)^2 - \mu_0 HM\cos\theta + K_1 \sin^2\theta \qquad (1)$$

Equilibrium is found by minimising $E(\theta)$

$$dE/d\theta = -\mu_0 M^2 \cos\theta \sin\theta + \mu_0 HM\sin\theta + 2K_1 \sin\theta \cos\theta = 0$$

Hence
$$-\mu_0 M\cos\theta + \mu_0 H + (2K_1/M)\cos\theta = 0$$

If $K_1 = 0$, $H = M\cos\theta$, otherwise $H = (1 - 2K_1/\mu_0 M^2)M\cos\theta$.

Since $H_s = 0.85 M_s$, $2K_1/\mu_0 M_s^2 = 0.15$. If $M_s = 500$ kAm$^{-1}$, for example, $K_1 = 24$ kJm$^{-3}$. Hence the intrinsic Co contribution to the anisotropy will be sufficient to ensure out-of-plane magnetization in a-$R_{20}Co_{80}$ films near compensation, when $M_s < 195$ kAm$^{-1}$. The origin of the intrinsic contribution could be interface anisotropy in the thin cobalt films, or a tendency towards preferential alignment of atom pairs relative to the film normal.

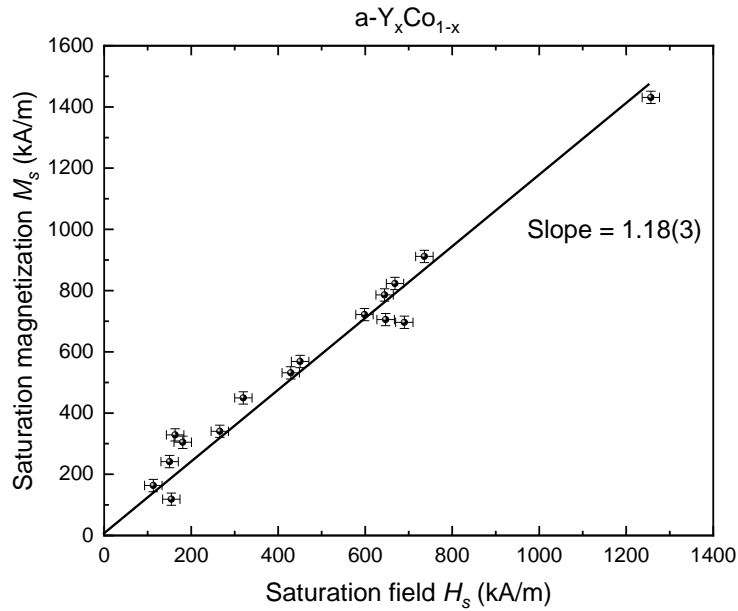

Figure 8. Plot of saturation magnetization versus saturation field. The slope > 1 indicates that intrinsic perpendicular anisotropy is overcome by shape anisotropy in the amorphous films.

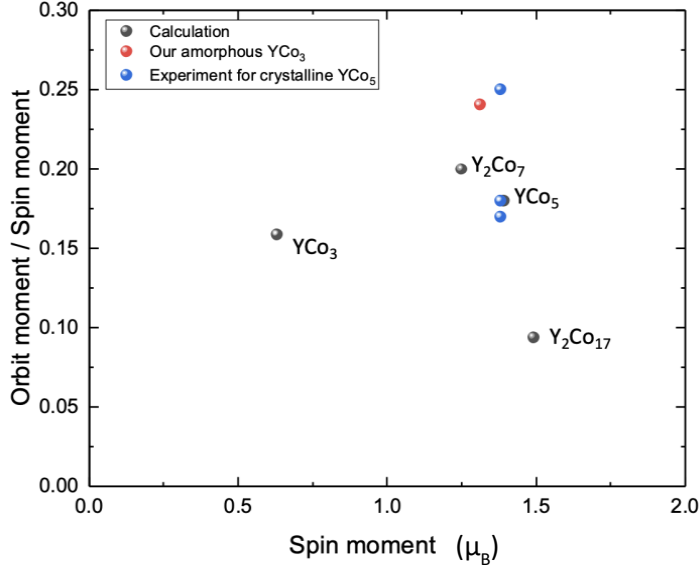

Figure 9 Ratio of average orbital to spin moment plotted against spin moment in crystalline Y-Co compounds.

Figure 9 compares the calculated average orbital moments for the crystalline Y-Co intermetallic compounds including both spin-orbit interaction and an orbital polarization term [23] with those measured for YCo$_5$ [24, 25] and that of a-YCo$_3$ deduced from XMCD (Fig. 5). YCo$_5$ is the intermetallic of cobalt with a nonmagnetic element that has biggest uniaxial anisotropy constant, $K_1$ = 6.5 MJm$^{-3}$, so the *local* anisotropy of cobalt in a-YCo$_3$ is expected to be similar in magnitude. In fact, Almeida *et al* inferred a value of 4 MJm$^{-3}$ from transverse-biased initial susceptibility measurements of sputtered amorphous films of a-Y$_x$Co$_{1-x}$ [16]. The large random anisotropy of the amorphous cobalt might have been expected to destroy collinear ferromagnetism in the a-Y$_x$Co$_{1-x}$ films, which is obviously not the case. This can be explained by the exchange averaging of the anisotropy, which arises when the ratio of exchange to anisotropy energy is sufficiently large. In the HPZ model of an amorphous ferromagnet [26], the local anisotropy is represented by a term $\bm{D}_i S_{zi}^2$ where the magnitude of $\bm{D}_i$ is constant, but its direction varies from site to site. Taking S = 1 for Co, the magnitude of $D$ corresponding to $K_1$ = 3.2 MJm$^{-3}$ is 5.6 K. $K_1$ is deduced from from the fact that the orbital moment of cobalt in a-YCo$_3$ is practically the same as that in YCo$_5$ and scaling by the number of cobalt atoms per unit volume. The Curie temperature was measured experimentally [27] and estimated theoretically [14] to be ≈ 760 K for a-Y$_{0.25}$Co$_{0.75}$. The corresponding molecular field [6]

$$\mu_0 H_i = 3k_B T_C / g(S + 1)\mu_B \qquad (2)$$

is 855 T, assuming g = 2 and S = 1. Therefore the ratio α of anisotropy to exchange energy per cobalt atom (5.6/1140) is of order 1/200. The ferromagnetic correlation length $L$ of can be estimated for a wandering axis ferromagnet by minimizing the sum of anisotropy and exchange energy. The number of atoms $N$ in the ferromagnetically-correlated volume is $(L/a)^3$ and the average energy per atom is the sum of anisotropy and exchange terms

$$E = -\sqrt{N} D S_x^2 / N + ZJS^2(\pi a/2L)^2 \qquad (3)$$

The first term describes the random single-ion anisotropy, and the second the increase in exchange due to misalignment of the spin of an atom with its Z neighbours by an angle $(\pi/2)/N^{1/3}$. Minimizing the energy with respect to $L$ gives [6]

$$L = (1/9\alpha^2)\pi^4 a \qquad (4)$$

Taking $1/\alpha = 200$ and $a = 0.28$ nm for a-YCo$_3$, we estimate $L = 120$ μm, which is comparable with the domain size in a normal ferromagnet.

When the argument is applied to the rare earth sublattice in a-R-Co alloys where R is a heavy rare earth between Tb and Tm, the anisotropy per rare-earth is greater in magnitude, but the R-Co exchange is only a few percent of the Co-Co exchange, leading to a random, noncollinear structure of the rare earth subnetwork, to be discussed in a companion paper. Compositions lying between a-RCo$_2$ and a-RCo$_3$, $0.25 < x < 0.33$, when R = Gd, [28-32] Tb [28, 33-35] Dy [8,35-38] Ho [29,36] and Er [28,38] may exhibit compensation.
.

The appearance of magnetism in bulk, crystalline Y$_x$Co$_{1-x}$ material occurs just below x = 0.33; the YCo$_2$ laves phase, which has cobalt in 3:6 Y:Co coordination is a band metamagnet with an enhanced Pauli susceptibility [20]. Three yttrium neighbours suffice to destroy the cobalt moment. The moment is more robust in the amorphous state, due in part to the lower density, reducing 4$d$-3$d$ charge transfer and hybridization [14, 39], but the critical number of Y neighbours is again close to three, which is reached at a greater value of $x_0$. Assuming the amorphous alloys remain strong ferromagnets, the critical concentration $x_c$ for the appearance of magnetism in the magnetic valence model [6] should be 0.40, assuming Co has $N_{4s} = 0.6$ unpolarized 4$s$ electrons, The electron counting however takes no account of the *orbital* moment that we find in the amorphous films. Furthermore, we are measuring thin films, not bulk material. The surfaces in crystalline YCo$_2$, layers remain ferromagnetic [40], and bulk defects are magnetic [41]. Interface layers in our amorphous thin films account for 6-7% of the film thickness, and could retain their magnetism for longer than the rest of the film. Some yttrium 4$d$ polarization is also possible. Yttrium moments of -0.2 to -0.4 μ$_B$ are found in the band calculations for Y-Co intermetallics [23], although they have not been detected in XMCD [42].

It is interesting to compare the strongly ferromagnetic cobalt subnetwork in a-Y$_x$Co$_{1-x}$ with the noncollinear magnetism resulting from a distribution of ferromagnetic and antiferromagnetic Fe-Fe exchange interactions in a-Y$_x$Fe$_{1-x}$ [43]. The critical concentration for the appearance of magnetism in the amorphous iron alloys greater than that for cobalt, $x_0 \approx 0.6$. The magnetization curves are nonlinear and cannot be saturated — the aligned moment per iron in Y$_{12}$Fe$_{88}$ in 5 T at 4 K is 1.51 μ$_B$ yet the average magnitude of the iron moment deduced from the hyperfine field in 0 T is 2.08 μ$_B$ [44]. The magnetic structure is asperomagnetic, where the spins are frozen in random directions with short-range ferromagnetic correlations below a magnetic ordering temperature of 109 K. This reflects the sensitivity of Fe-Fe exchange to the interatomic distance. Close-packed fcc ion, with a nearest-neighbour distance of 254 pm is ferromagnetic, but shorter interatomic distances in the amorphous close-packed structure introduce antiferromagnetic interactions into the nearest-neighbour exchange distribution, which is centred at a positive value. However, a small expansion of the amorphous structure by absorbed hydrogen is sufficient to produce colinear ferromagnetism with a full iron moment of 2.3 μ$_B$ and $T_C$ = 500 K [45].

The scaling for the normal and spontaneous components of the Hall effect and magnetoresistance illustrated in Fig. 7 are directly related to the behavior of the average, macroscopic magnetization which rotates coherently in fields below 1T. The antisymmetrized spontaneous Hall voltage $V_{xy}^{as} \propto M_z$ and $M_z \propto H$ for fields below saturation. The symmetrized resistive pickup $V_{yy}$ is proportional to the in-plane magnetization component $M_y^2$, which varies as $\cos^2[\sin^{-1}(H/2K_1)]$. At fields above saturation the dependences on field are dominated by the ordinary Hall effect, linear in $H$, and the conventional band magnetoresistance, quadratic in $H$. The microscopic distribution in magnitude and direction of the local cobalt anisotropy is modelled using a gaussian distribution with a width of ≈18%. Examples of the fits were shown, together with the residuals, on Fig. 6.. The large observed magnitude of the normal Hall effect indicates a low density of carriers with poor mobility at the Fermi level. There is no change of sign of the spontaneous Hall component. The same model of uniform rotation is used to interpret the ordinary band and anomalous magnetoresistance on the four panels of Fig 6. Discrepancies are visible only for $x = 045$, where the para process associated with the appearance of magnetism is beginning to contribute.

## 7 Conclusions

The ferromagnetism of cobalt in these amorphous thin films is strong and persistent. The cobalt moment disappears only at $x_0 \approx 0.50$, where an average Co atom in a random dense-packed structure is coordinated by 3.2 Y and 3.2 Co atoms. Three yttrium nearest neighbors destroy the moment of cobalt in either the crystalline or amorphous state. The packing fraction estimated from the model structure is 0.63 for all compositions, but experimental values tend towards 0.74 as $x \Rightarrow 0.5$.

A large orbital moment on cobalt, amounting to 0.32 $\mu_B$ in a-YCo$_3$, is associated with the low densities of the Co-rich compositions. The resulting local anisotropy of cobalt is as strong as it as that in YCo$_5$, the crystalline intermetallic of cobalt with the strongest uniaxial cobalt anisotropy, but exchange averaging associated with the high Curie temperature of 760 K [14, 27] ensures that deviations from collinear ferromagnetism and coherent rotation of the macroscopic magnetization are negligible. It is noteworthy that a system with such strong local anisotropy exhibits no coercivity or hysteresis and behaves uniformly on a macroscopic scale because of the strong Co-Co exchange. The amorphous alloys with $x \approx 0.15$ are expected to exhibit even larger orbital moments, with an extrapolated Curie temperatures that surpasses that of crystalline cobalt.

Although the magnetization of the amorphous thin films lies in-plane, there is an intrinsic perpendicular component of the anisotropy that is overcome by shape anisotropy throughout the ferromagnetic yttrium-cobalt series. This intrinsic term would, however, be sufficient to induce perpendicular magnetic anisotropy in other a-R-Co alloys with a much lower net magnetization that are close to ferrimagnetic compensation.

The present study of the cobalt subnetwork in alloys with a nonmagnetic rare earth will enable a more accurate description of the noncollinear sperimagnetic structures which arise in amorphous rare-earth cobalt alloys with heavy rare earths other than Gd and exhibit compensation near room temperature.

**Acknowledgements** This work was supported by Science Foundation Ireland under grants 16/IA/4534 ZEMS, 12/RC/2278 AMBER, 17/NSFC/5294 MANIAC and EU FET Open grant 737038 TRANSPIRE. ZH acknowledges support of a Postgraduate Scholarship from Trinity College Dublin.